\documentclass[twoside]{LCWS11}
\usepackage[latin1]{inputenc}
\usepackage[dvips]{graphicx,epsfig,color}
\usepackage{wrapfig,rotating}
\usepackage{amssymb,amsmath,array}
\usepackage{cite}
\usepackage{slashed}

\newcommand{\nn}{\nonumber}
\newcommand{\Mpl}{\overline{M}_{\rm Pl}}
\newcommand{\gld}{\tilde G}
\newcommand{\no}{\tilde{\chi}^0_1}
\newcommand{\se}{\tilde e}

\begin{document}
\title{
Associated Production of Light Gravitinos\\ at Future Linear Colliders} 
\author{Kentarou Mawatari 
\thanks{This work has been supported in part by the Concerted Research
action ``Supersymmetric models and their signatures at the Large Hadron
Collider'' of the Vrije Universiteit Brussel and by the Belgian Federal
Science Policy Office through the Interuniversity Attraction Pole IAP
VI/11.} 
\vspace{.3cm}\\
Theoretische Natuurkunde and IIHE/ELEM, Vrije Universiteit Brussel,\\
and International Solvay Institutes, Pleinlaan 2, B-1050 Brussels, Belgium
}

\maketitle

\begin{abstract}
 We study light gravitino productions in association with a neutralino 
 at future linear colliders in a scenario in which the lightest
 SUSY particle is a gravitino and the produced neutralino
 promptly decays into a photon and a gravitino. Comparing with the
 multiple goldstino scenario, we show that energy and angular
 distributions of the photon provide valuable information on the SUSY
 masses as well as the SUSY breaking.  
\end{abstract}

\section{Introduction}

Gravitino productions in association with a SUSY particle are known
processes which become significant at colliders when the gravitino is
very light as $m_{3/2}\sim\cal O$($10^{-2}$~eV) or less, since the cross
sections are inversely proportional to the square of the Planck scale
times the gravitino mass  
\begin{align}
 \sigma\propto 1/(M_{\rm Pl\,}m_{3/2})^2.
\end{align}
When the associated SUSY particle is the next-to-lightest SUSY
particle (NLSP) and promptly decays into a SM particle and a LSP
gravitino, the production processes lead to particular collider
signatures, e.g. $\gamma+\slashed E$ and $j+\slashed E$, where
the missing energy is carried away by two gravitinos, and these signals 
set mass bounds on the gravitino and the other SUSY particles.
It should be noted that the gravitino mass is related to the SUSY
breaking scale as well as the Planck scale like
\begin{align}
 m_{3/2}\sim (M_{\rm SUSY})^2/M_{\rm Pl}.
\end{align}
The current experimental bound on the gravitino mass
from the single-photon plus missing-energy signal in
neutralino-gravitino associated productions is given by the LEP
experiment as a function of the neutralino and selectron masses, e.g.  
\begin{align}
 m_{3/2}\gtrsim 10^{-5}\ {\rm eV,}\
 {\rm i.e.}\ M_{\rm SUSY}\gtrsim 200\ {\rm GeV}, 
\end{align}
for $m_{\no}=140$ GeV and $m_{\se}=150$ GeV~\cite{Abdallah:2003np}.

Several theoretical studies on the $\no$-$\gld$ productions in $e^+e^-$  
collisions had been done before especially for the
LEP~\cite{Fayet:1986zc,Dicus:1990vm,Lopez:1996gd}, and recently the
process was restudied for future linear colliders with the then current
simulation tools~\cite{Hagiwara:2010pi,Mawatari:2011jy} in
Ref.~\cite{Mawatari:2011cu}. 
We note that such a very light gravitino is suggested by the context of
no-scale supergravity~\cite{Ellis:1984kd,Lopez:1992ni} and some
extra-dimensional models~\cite{Gherghetta:2000qt}, while in typical
gauge-mediated SUSY breaking (GMSB) scenarios we expect a mass of
1~eV--10~keV~\cite{Giudice:1998bp}.  

In this report, we extend our previous study on the process
$e^+e^-\to\no\gld$~\cite{Mawatari:2011cu} with the latest tools, 
{\tt FeynRules}~\cite{Christensen:2008py,Degrande:2011ua} and
{\tt MadGraph5}~\cite{Alwall:2011uj}, and make a comparison with the
multiple goldstino scenario, which was presented recently in
Ref.~\cite{Argurio:2011gu}.

\section{Goldstini in gauge mediation}

Multiple goldstino models, so-called {\it goldstini}
models~\cite{Cheung:2010mc}, in the framework of gauge mediation are 
characterized by a visible sector (e.g.~the MSSM) coupled by gauge
interactions to more than one SUSY breaking
sector~\cite{Argurio:2011hs}. The spectrum consists of a light gravitino
LSP, behaving as a goldstino, and a number of neutral fermions (the
pseudo-goldstini) with a mass between that of the LSP and that of the
lightest observable-sector SUSY particle (LOSP). Here we consider a
situation where the LOSP is the lightest neutralino and there is only
one pseudo-goldstino with a mass of ${\cal O}(100)$~GeV. The coupling of 
the MSSM particles to the pseudo-goldstino can be enhanced with respect
to those of the gravitino giving rise to characteristic signatures. The
relevant pseudo-goldstino interaction Lagrangian is shown in
Appendix~A. 

\begin{figure}
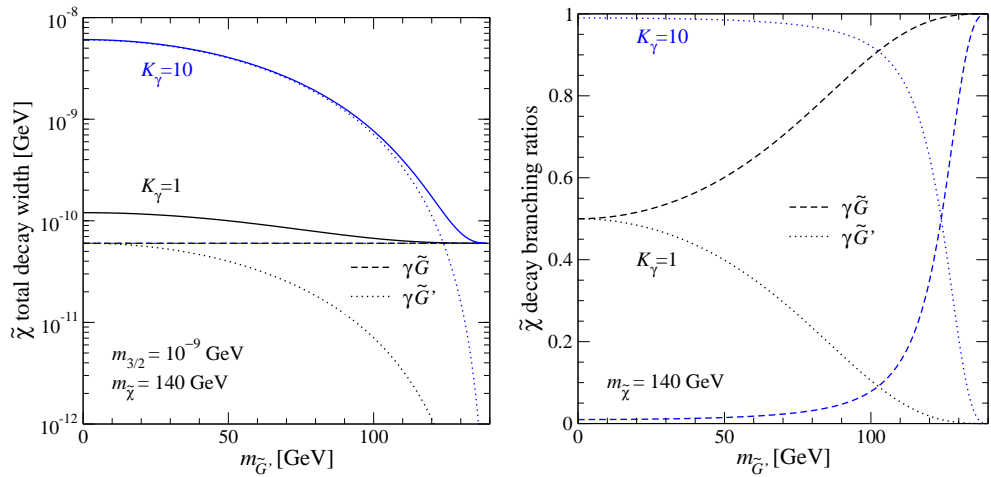

\begin{center}
 \includegraphics[height=6.25cm,clip]{photinodecay_t.eps}\quad
 \includegraphics[height=6.25cm,clip]{photinodecay_b.eps}
 \caption{Total decay width (left) and decay branching ratios (right) of 
 the lightest neutralino, assumed as a pure photino, as a function of
 the pseudo-goldstino mass for $K_{\gamma}=1$ (black lines) and 10 (blue
 lines).} 
\label{fig:decay}
\end{center}
\end{figure} 

To highlight the differences with respect to the case of the single SUSY
breaking sector and the role played by the extra parameters $K$,
characterizing the pseudo-goldstino couplings, we present the total
decay width and the decay branching ratios of the lightest neutralino in
Fig.~\ref{fig:decay}. For simplicity we assume the neutralino is a pure
photino in this report. The partial decay width for the decay into a
photon and a pseudo-goldstino is given by~\cite{Argurio:2011gu} 
\begin{align}
 \Gamma(\no\to\gamma\gld') = 
  \frac{K_{\gamma}^2|C_{\gamma\tilde\chi_1}|^2 m_{\no}^5}
       {48\pi\Mpl^2m_{3/2}^2}
  \bigg(1-\frac{m_{\gld'}^2}{m_{\no}^2}\bigg)^3
\label{n1decay}
\end{align}
with the reduced Planck mass 
$\Mpl\equiv M_{\rm Pl}/\sqrt{8\pi}\sim2.4\times10^{18}$~GeV and the mass 
of the gravitino (i.e. the true goldstino) $m_{3/2}$.
$C_{\gamma\tilde\chi_1}$ is defined in Appendix~A and equal to unity for 
the photino case. The $m_{\gld'}=0$ limit with $K_{\gamma}=1$
reduces~\eqref{n1decay} to $\Gamma(\no\to\gamma\gld)$. The decay width
and the branching ratios strongly depend on the pseudo-goldstino mass
and the $K_{\gamma}$ factor. We refer to~\cite{Argurio:2011gu} for more
details and the $\no\to Z\gld'$ decay.

\section{Single-photon plus missing energy signal}

\begin{figure}
\begin{center}
 \includegraphics[width=0.425\columnwidth,clip]{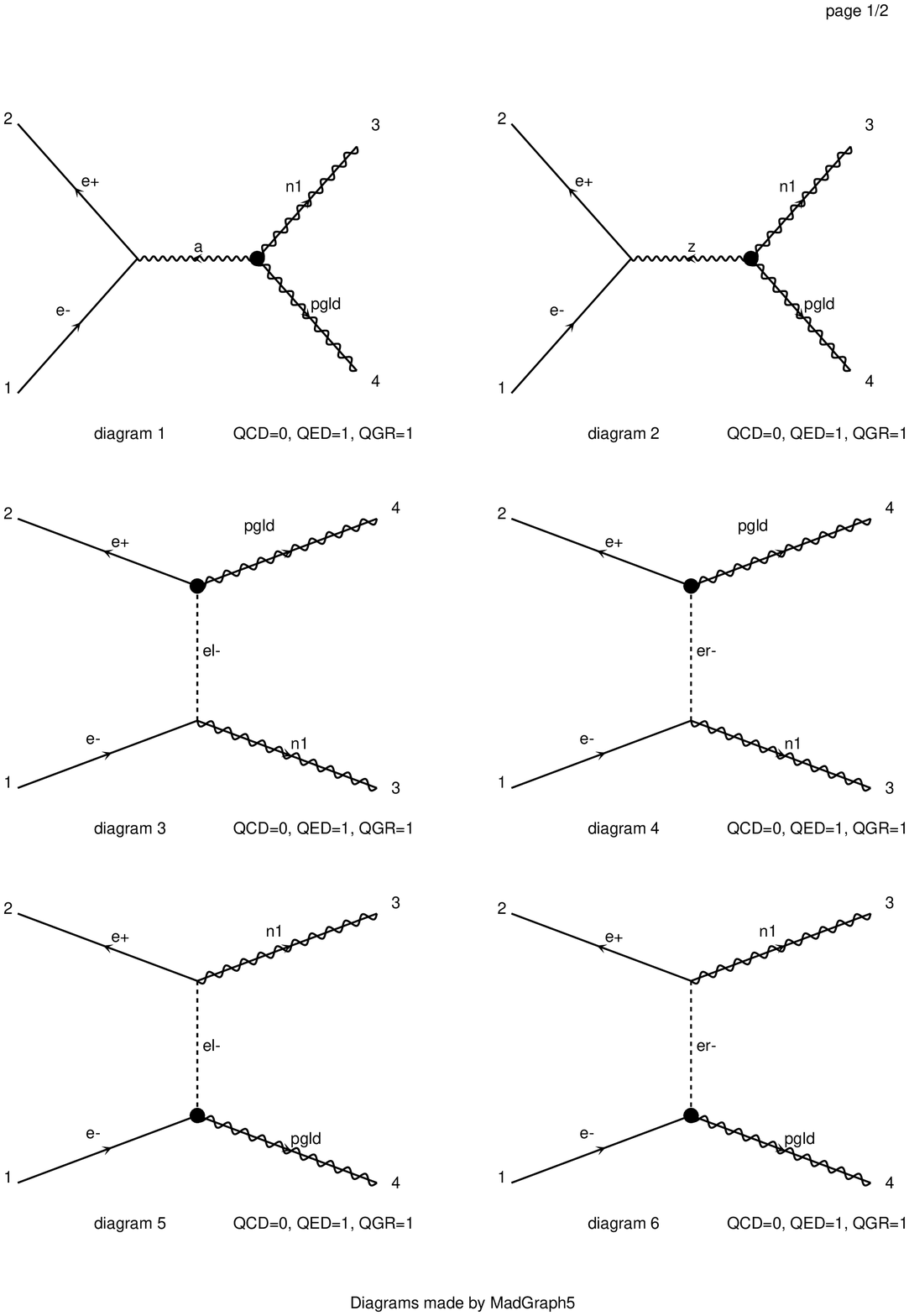}\quad
 \includegraphics[width=0.525\columnwidth,clip]{xsec_mpgld.eps}
 \caption{(Left) Feynman diagrams for the process $e^+e^-\to\no\gld'$, 
  generated by {\tt MadGraph5}~\cite{Alwall:2011uj}.
  (Right) The total cross sections at $\sqrt{s}=500$~GeV (black lines)
 and 1~TeV (red lines) as a function of the pseudo-goldstino mass, for
 various values of $K_{\gamma}$ and $K_{e}$.} 
\label{fig:xsec}
\end{center}
\end{figure}

As mentioned before, in typical GMSB models, the gravitino mass is not
accessible in colliders since the associated production cross section is
too small. In the case of the pseudo-goldstino, however, the cross
section can be enhanced by the coupling factors $K$ while keeping the
gravitino mass as $m_{3/2}\sim$~eV, i.e. $M_{\rm SUSY}\sim 100$~TeV.

In this report, we consider pseudo-goldstino productions in association
with a neutralino in $e^+e^-$ collisions, where the produced LOSP
neutralino subsequently decays into a photon and a (almost massless)
gravitino or into a photon and a (massive) pseudo-goldstino,
\begin{align}
 e^+e^-\to\no\gld';\quad \no\to\gamma\gld\ {\rm or}\ 
 \gamma\gld'.
\label{process}
\end{align}  
The decay fraction is determined by $m_{\gld'}$ and $K_{\gamma}$ as one
can see in Fig.~\ref{fig:decay}.
Feynman diagrams for the production process are shown in
Fig.~\ref{fig:xsec} (left). Since the neutralino is assumed here to be a
pure photino, we can neglect the diagram 2. In the $t$- and $u$-channels
the intermediate particle is either the left- or right-handed selectron,
and we assume that the coupling factor $K_{e}$ is the same for both
selectrons; see also the interaction Lagrangian in Appendix~A.  
All the helicity amplitudes for the production process are presented in
the $m_{\gld'}=0$ limit in~\cite{Mawatari:2011cu}, while the spin summed
amplitude squared is shown in~\cite{Argurio:2011gu}.  

Figure~\ref{fig:xsec} (right) shows the production cross sections as a
function of the pseudo-goldstino mass for some values of the parameters 
$K_{\gamma}$ and $K_{e}$. Here we take the masses as
$m_{3/2}=10^{-9}$~GeV, $m_{\no}=140$~GeV and
$m_{\tilde e_L}=m_{\tilde e_R}=400$~GeV, while those masses as well as
beam polarizations can change the cross section~\cite{Mawatari:2011cu}.  
It should be stressed that the cross section scales with
$K_{\gamma,e}^2/m_{3/2}^2$, and hence the cross section in the 
$m_{\gld'}=0$ limit for $m_{3/2}=10^{-9}$~GeV with
$K_{\gamma}=K_{e}=10^4$ is equal to that for $m_{3/2}=10^{-13}$~GeV in
the single sector scenario.
There is a destructive interference between the diagrams, and thus the
cross section for large $K_e$ turns out to be greater than the cross
section when both $K_{\gamma}$ and $K_e$ are large. We notice that
rather large values of $K_{\gamma}$ and $K_e$ are required to obtain the
cross section around ${\cal O}(10^{2-3})$~fb with the eV order gravitino
mass, while such large values are not favored by the stability of the
SUSY breaking vacuum~\cite{Argurio:2011gu}.  

\begin{figure}
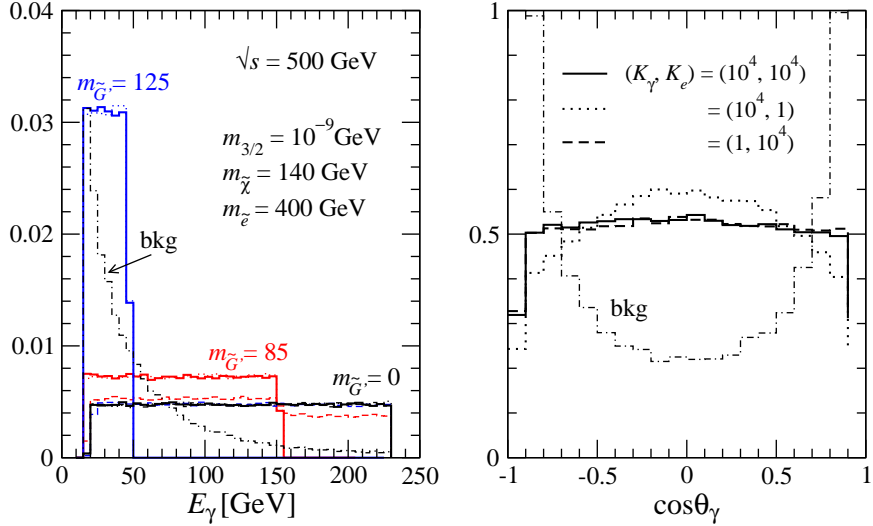

\begin{center}
 \includegraphics[height=7cm,clip]{monophoton_e.eps}\quad
 \includegraphics[height=7cm,clip]{monophoton_cos.eps}
 \caption{Normalized energy (left) and angular (right) distributions of 
 the photon for 
 $e^+e^-\to\no\gld'\to\gamma\gld^{(')}\gld'$ at $\sqrt{s}=500$ GeV.}
\label{fig:photon}
\end{center}
\end{figure} 

Since the $\no\to\gamma\gld^{(')}$ decay is isotropic, the photon
distribution is given by purely kinematical effects of the decaying
neutralino. Figure~\ref{fig:photon} shows normalized energy (left) and
angular (right) distributions of the photon for the
signal~\eqref{process} as well as the SM background at 
$\sqrt{s}=500$~GeV. The minimal cuts for the detection of photons 
\begin{align}
 E_{\gamma}>0.03\,\sqrt{s}=15\ {\rm GeV},\quad |\eta_{\gamma}|<2,
\end{align}
and the $Z$-peak cut to remove the SM $(Z\to\nu\bar\nu)\gamma$
background 
\begin{align}
 E_{\gamma}<\frac{s-m_Z^2}{2\sqrt{s}}-5\Gamma_Z\sim 230\ {\rm GeV},
\end{align}
are imposed.
The most significant background coming from the $t$-channel
$W$-exchange process can be reduced by using polarized $e^{\pm}$ beams, 
while the distributions do not change so much both for signal and
background; see more quantitative details in~\cite{Mawatari:2011cu}.

The energy distributions are flat, and the maximal and minimal energy
are given by 
\begin{align}
 E^{\mathrm{max},\mathrm{min}}_\gamma=
 \frac{\sqrt{s}}{4}\hat\beta 
 \bigg(1+\frac{m_{\no}^2-m_{\gld'}^2}{s}\pm\beta\bigg),
\end{align}
where $\hat\beta=1$ for $\no\to\gamma\gld$ and 
$(1-\frac{m_{\gld'}^2}{m_{\tilde\chi}^2})$ for $\no\to\gamma\gld'$, and   
$\beta=\bar\beta(\frac{m_{\tilde\chi}^2}{s},\frac{m_{\gld'}^2}{s})$
with $\bar\beta(a,b)=(1+a^2+b^2-2a-2b-2ab)^{1/2}$. 
The higher edge can determine the pseudo-goldstino mass when the 
$\no\to\gamma\gld'$ decay is significant. 

On the other hand, the angular distributions are not sensitive to the
pseudo-goldstino mass, and hence only the $m_{\gld'}=0$ case is shown in 
Fig.~\ref{fig:photon} (right). Instead the distributions depend on the
coupling factors $K_{\gamma}$ and $K_e$. 
This is because that the distributions are determined by how much
the $s$-channel and $t,u$-channel diagrams contribute.
The selectron-exchange contribution can be enhanced by increasing $K_e$
as well as $m_{\se}$, which lead to the flatter distributions; see more
details on the $m_{\se}$ dependence in~\cite{Mawatari:2011cu}.

\section{Summary}

We extended our previous study on the mono-photon plus missing energy
signal in the $\no$-$\gld$ associated production at future linear
colliders~\cite{Mawatari:2011cu}. All the results presented here can be
obtained numerically running {\tt MadGraph5}~\cite{Alwall:2011uj}
simulations adapted to the (pseudo)-goldstino scenario (building
on~\cite{Mawatari:2011jy}), having implemented the model using 
{\tt FeynRules}~\cite{Christensen:2008py,Degrande:2011ua}. Comparing
with the multiple goldstino scenario~\cite{Argurio:2011gu}, we showed
that the energy and angular distributions of the photon can explore the
SUSY masses as well as the SUSY breaking mechanism. 

Before closing, we note that single-electron plus missing energy
signals in $\se$-$\gld$ associated productions at $e\gamma$ colliders
were also studied in detail in Ref.~\cite{Mawatari:2011cu}.

\section*{Acknowledgments}

I thank B.~Oexl and Y.~Takaesu for the collaboration. I also wish to
thank R.~Argurio, K.~De~Causmaecker, G.~Ferretti and A.~Mariotti for
useful comments on the goldstini part.

\section*{Appendix A: Pseudo-goldstino interaction Lagrangian}

We briefly present the relevant terms of the pseudo-goldstino
interaction Lagrangian for our study both in the derivative and
non-derivative forms, which were implemented into
{\tt FeynRules}~\cite{Christensen:2008py} to obtain the {\tt UFO} model
file~\cite{Degrande:2011ua} for {\tt MadGraph5}~\cite{Alwall:2011uj}. 

\begin{itemize}
\item In the derivative form:
\begin{align}
 {\cal L}_{\partial\gld'}
 =&\pm\frac{iK_{e}}{\sqrt{3}\,\Mpl\,m_{3/2}}
   \big[
   \partial_{\mu}\bar{\psi}_{\gld'}\gamma^{\nu}\gamma^{\mu}P_{\pm}\psi_e^{}\,
     \partial_{\nu}\phi_{\se_{\pm}}^*
  -\bar{\psi}_e^{}P_{\mp}\gamma^{\mu}\gamma^{\nu}\partial_{\mu}\psi_{\gld'}\,
     \partial_{\nu}\phi_{\se_{\pm}}^{} 
   \big] \nn\\
 &-\frac{iK_{\gamma}C_{\gamma\tilde{\chi}_i}}
       {4\sqrt{6}\,\Mpl\,m_{3/2}}
  \partial_{\mu}\bar{\psi}_{\gld'}[\gamma^{\nu},\gamma^{\rho}]\gamma^{\mu}
    \psi_{\tilde{\chi}^0_i}
  (\partial_{\nu}A_{\rho}-\partial_{\rho}A_{\nu}) \nn\\
 &-\frac{iK_{Z_T}C_{Z_T\tilde{\chi}_i}}{4\sqrt{6}\,\Mpl\,m_{3/2}}
  \partial_{\mu}\bar{\psi}_{\gld'}
    [\gamma^{\nu},\gamma^{\rho}]\gamma^{\mu}\psi_{\tilde{\chi}^0_i}
  (\partial_{\nu}Z_{\rho}-\partial_{\rho}Z_{\nu}) \nn\\
 &-\frac{2m_ZK_{Z_L}C_{Z_L\tilde{\chi}_i}}
       {\sqrt{6}\,\Mpl\,m_{3/2}}
  \partial_{\mu}\bar{\psi}_{\gld'}\psi_{\tilde{\chi}^0_i}Z^{\mu},
\label{L_int}
\end{align}
where the couplings related to the neutralino mixing defined by
$X_i=U_{ij}\tilde\chi^0_j$ in the
$X=(\tilde B,\tilde W^3,\tilde H^0_d,\tilde H^0_u)$ basis are  
\begin{align}
 C_{\gamma\tilde\chi_i}&=U_{1i}\cos{\theta_W}+U_{2i}\sin{\theta_W},\nn\\
 C_{Z_T\tilde\chi_i}&=-U_{1i}\sin{\theta_W}+U_{2i}\cos{\theta_W}, \nn\\
 C_{Z_L\tilde\chi_i}&= U_{3i}\cos{\beta}-U_{4i}\sin{\beta},
\label{couplings}
\end{align}
with the ratio of the vacuum expectation value of the two Higgs doublets
$\tan\beta$. $K_{e}$, $K_{\gamma}$, $K_{Z_T}$ and $K_{Z_L}$ are
parameters characterizing the pseudo-goldstino
couplings~\cite{Argurio:2011gu}, and the $K=1$ limit  
reduces~\eqref{L_int} to that for the pure goldstino. 
  
\item In the non-derivative form:
\begin{align}
 {\cal L}_{\eth\gld'}
 =&\mp\frac{iK_{e}m_{\se_{\pm}}^2}{\sqrt{3}\,\Mpl\,m_{3/2}}
   \big[ \bar{\psi}_{\gld'}P_{\pm}\psi_e^{}\,\phi_{\se_{\pm}}^*
        -\bar{\psi}_e^{}P_{\mp}\psi_{\gld'}\,\phi_{\se_{\pm}}^{} 
   \big] \nn\\
 &-\frac{K_{\gamma}C_{\gamma\tilde{\chi}_i}m_{\tilde{\chi}^0_i}}
       {4\sqrt{6}\,\Mpl\,m_{3/2}}
  \bar{\psi}_{\gld'}[\gamma^{\mu},\gamma^{\nu}]\psi_{\tilde{\chi}^0_i}
  (\partial_{\mu}A_{\nu}-\partial_{\nu}A_{\mu}) \nn\\
 &-\frac{ K_{Z_T}C_{Z_T\tilde{\chi}_i}m_{\tilde{\chi}^0_i}
         +K_{Z_L}C_{Z_L\tilde{\chi}_i}m_{Z}}
       {4\sqrt{6}\,\Mpl\,m_{3/2}}
  \bar{\psi}_{\gld'}[\gamma^{\mu},\gamma^{\nu}]\psi_{\tilde{\chi}^0_i}
  (\partial_{\mu}Z_{\nu}-\partial_{\nu}Z_{\mu}) \nn\\
 &-\frac{im_Z( K_{Z_T}C_{Z_T\tilde{\chi}_i}m_{Z}
         +K_{Z_L}C_{Z_L\tilde{\chi}_i}m_{\tilde{\chi}^0_i})}
       {\sqrt{6}\,\Mpl\,m_{3/2}}
  \bar{\psi}_{\gld'}\gamma^{\mu}\psi_{\tilde{\chi}^0_i}Z_{\mu}.
\end{align}
\end{itemize}


\begin{footnotesize}


\end{footnotesize}


\end{document}